\documentclass{article}

\usepackage[square,sort,comma,numbers]{natbib}

\usepackage[preprint]{neurips_2023}

\usepackage[utf8]{inputenc} 
\usepackage[T1]{fontenc}    
\usepackage{hyperref}       
\usepackage{url}            
\usepackage{booktabs}       
\usepackage{amsfonts}       
\usepackage{nicefrac}       
\usepackage{microtype}      
\usepackage{xcolor}         
\usepackage{graphicx}

\title{Musical Form Generation}

\author{Lilac Atassi\\
  Department of Music\\
  University of California, San Diego\\
  \texttt{latassi@ucsd.edu} \\
}

\begin{document}

\maketitle

\begin{abstract}
  While recent generative models can produce engaging music, their utility is limited. The variation in the music is often left to chance, resulting in compositions that lack structure. Pieces extending beyond a minute can become incoherent or repetitive. This paper introduces an approach for generating structured, arbitrarily long musical pieces. Central to this approach is the creation of musical segments using a conditional generative model, with transitions between these segments. The generation of prompts that determine the high-level composition is distinct from the creation of finer, lower-level details. A large language model is then used to suggest the musical form.
\end{abstract}

\section{Introduction}

Music generative models have recently demonstrated their ability to produce remarkable musical compositions, potentially attributed to their proficiency as sequence decision-makers \cite{conditional}  However, a significant hindrance to their practical utility lies in their deficiency of composition-level structure. This becomes evident after approximately one minute of generated music. Neither a musician nor a novice user could use them to create a long, appealing piece of music."

End-to-end music generative models often focus more on small-scale structures, such as chords and phrases, over large-scale structures like musical form. This emphasis can hinder a machine learning model's ability to innovate at the composition level. Segmenting music into dual structural levels can improve control over its form. The concept of hierarchical music models dates back at least to the work of Lerdahl and Jackendoff \cite{lerdahl1996generative}. Conditional generative models pave the way to distinguish between controlling composition-level music and managing small-scale structures. By combining music segments generated from distinct prompts, one can craft an extended piece. The prompt sequence dictates the form, and each individual prompt guides the small-scale structures within its associated segment. This distinction allows the generative model to more abstractly conceptualize musical form.

Generative models introduce musical variation through sampling. Typically, a temperature parameter is fine-tuned, and the best results are often handpicked to minimize repetition and ensure coherence. However, instead of leaving the overarching structure to random chance, our method reduces the temperature parameter during sampling. We introduce composition-level variations via text prompts. Some samples demonstrating transitions between segments using our approach can be accessed at \href{https://lilac-code.github.io/compose.html}{\nolinkurl{lilac-code.github.io}}."

\section{Method}


\begin{figure}
    \centering
    \includegraphics[width=0.7\linewidth]{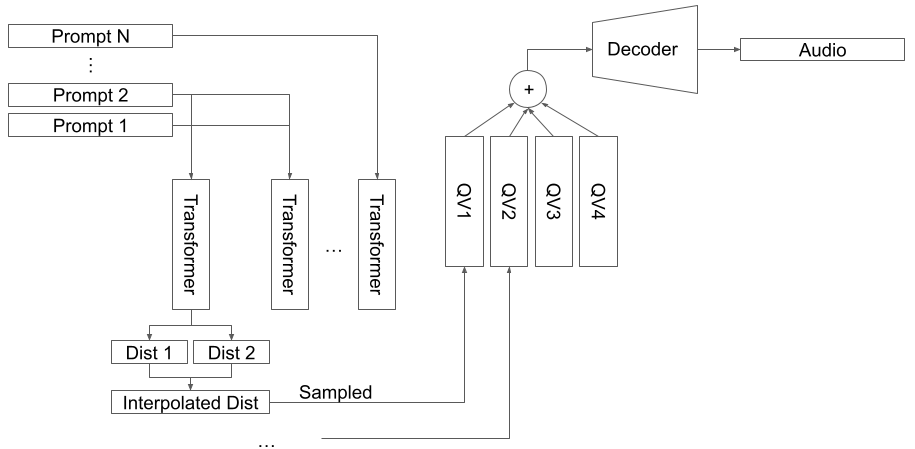}
    \caption{In our proposed method, the transformer model produces new tokens by sampling from a probability distribution. This distribution is a weighted blend of two distinct distributions, each conditioned on a text prompt. Over time, these weights adjust to facilitate a transition from one prompt to the next.}
    \label{fig:sys}
\end{figure}

Certain model architectures align more effectively with our proposed approach. AudioLDM 2 \cite{liu2023audioldm} employs a language model grounded on text prompts to produce tokens within the latent space of AudioMAE \cite{huang2022masked}. Following this, a latent diffusion model conditioned on these tokens generates a latent vector, which the VAE decoder then translates into a spectrogram. However, during our experiments, we observed that the latent diffusion model in this architecture struggled to produce transitions. This limitation primarily stems from the diffusion model's dependency on a robust classifier guidance scale to yield high-quality audio. Moreover, the conditioning occurs over a global timeframe. Consequently, when tasked with bridging two fixed endpoints to form a transition, the diffusion model either results in a noisy sound with a diminished guidance scale or a singular state when the guidance scale is heightened.

The streamlined architecture of MusicGen \cite{copet2023simple} offers enhanced control over the classifier-free guidance scale, which influences the generation of tokens. As a result, the MusicGen model can be readily adapted to our method. In MusicGen, a language model is trained to produce quantized latent vectors from a given prompt. These vectors are subsequently transformed into audio by the decoder of EnCodec \cite{defossez2022highfi}.


We run the transformer inference using two conditioning vectors, derived from two consecutive prompts. The next token is generated by sampling from the weighted combination of the two probability distributions. Since the transformer takes into account both past tokens and the two prompts when generating each token, the mentioned jarring jump issue is eliminated with this approach.

After obtaining the two probability distributions over all dictionary codes from the two prompts, we compute their weighted average. Fig.~\ref{fig:sys} illustrates this sampling process. We perform a linear interpolation between the two distributions over an interval of 5 seconds or less. In instances where the interval exceeds 5 seconds, our experiments showed a notable chance of audio artifacts appearing. During this interpolation interval, we also boost the temperature parameter by 50\%. When using beam search as the sampling method, we also double the top K parameter during the transition phase.
In our tests, depending on the distinctiveness of the prompts and their musical differences, fine-tuning these two parameters proved essential for achieving seamless transitions.

Our proposed method provides users with a dual level of control over
musical form. Users have the option to input a sequence of text
prompts, specifying the desired length for each musical segment.
Alternatively, they can employ a different system, such as a large
language model with in-context learning, to generate the sequence of
text prompts for the music generative model. During our experiments,
we observed that when transitioning between two prompts that represent
a significant musical contrast, for example, moving from electronic
dance music to classical music, the model faced challenges. It either
hesitated to switch its musical context or produced a decrease in the
quality of the generated audio.

\section{Conclusions}

Current text-to-music models are limited by their reliance on a single text prompt, though music generated under a minute can be engaging. This paper proposes a method that extends the use of such models, offering control over the musical form of longer pieces. Having high-level control over the musical form facilitates research into models that operate at the abstract composition level. This addresses a significant gap in the existing methods and tools for music generation.

\begin{ack}
I would like to express my gratitude for Shahrokh Yadegari, Robert Wannamaker, and Mosalam Ebrahimi whose feedback helped shape the ideas that led to the work presented in this paper. Additionally, I would like to thank NVIDIA Corporation for donating the graphics cards used in this research.
\end{ack}

\section*{Ethical Considerations}
The work showcased has ethical considerations in line with the
foundational generative models. As the public learn more about such models some of the risks are mitigated.


\bibliographystyle{ieeetr}
\bibliography{neurips_2023}







\end{document}